# Tunnel barrier enhanced voltage signals generated by magnetization precession of a single ferromagnetic layer

T. Moriyama,[1] R. Cao,[1] X. Fan,[1] G. Xuan,[2] B. K. Nikolić,[1] Y. Tserkovnyak,[3] J. Kolodzey,[2] and John Q. Xiao[1]

[1]*Department of Physics and Astronomy, University of Delaware, Newark, DE 19716, USA*

[2]*Department of Electrical and Computer Engineering, University of Delaware, Newark, DE 19716, USA*

[3]*Department of Physics and Astronomy, University of California, Los Angeles, CA 90095, USA*

**Abstract:**

We report the electrical detection of magnetization dynamics in an Al/AlO$_x$/Ni$_{80}$Fe$_{20}$/Cu tunnel junction, where a Ni$_{80}$Fe$_{20}$ ferromagnetic layer is brought into precession under the ferromagnetic resonance (FMR) conditions. The dc voltage generated across the junction by the precessing ferromagnet is enhanced about an *order of magnitude* compared to the voltage signal observed when the contacts in this type of multilayered structure are ohmic. We discuss the relation of this phenomenon to magnetic spin pumping and speculate on other possible underlying mechanisms responsible for the enhanced electrical signal.

In recent years, basic and applied research in metal-based spintronics has shifted increasingly from the static to the dynamic magnetic properties in hybrid nanostructures composed of ferromagnetic and normal metal layers [1-5]. A variety of experimentally observed phenomena involving nonlocal magnetization dynamics in magnetic multilayers are due to two complementary effects: (i) the transfer of spin angular momentum accompanying charge currents driven by the applied bias voltage between ferromagnetic layers results in torques that (for sufficiently high current densities) generate spontaneous magnetization precession and switching [1]; and (ii) the precessing magnetization of a ferromagnet (*FM*) pumps spins into adjacent normal metal layers (*NM*) with no applied bias [2, 5, 6]. In particular, the spin pumping effect [5] is a promising candidate for realizing a spin battery device [7] as a source of elusive *pure* spin currents (not accompanied by any net charge transport) emitted at the *FM/NM* interface, where steady magnetization precession of the *FM* layer is sustained by the absorption of external rf radiation under the FMR conditions. Another promising application of magnetization dynamics is microwave-assisted reduction of the switching field of *FM*, which could play an important role in advanced recording media [4].

Thus far, however, the spin pumping effect has been demonstrated mostly *indirectly* as an additional contribution to the FMR linewidth in *FM/NM* multilayers (where the *NM* is Pt, Pd, Cu, etc.) that can be described as the interface-induced enhancement of the Gilbert dumping constant [8-11]. The vigorous pursuit of direct electrical detection of spin pumping has led to theoretical proposals [6] to use a single precessing *FM* as *both the source and detector* of pumped spin accumulation in *NM* layers. This adroit scheme has been realized in a very recent experiment [2] measuring the difference in voltages of the order of several hundred nanovolts between two *FM/NM* interfaces of a *NM1/FM/NM2* trilayer.

In this Letter, we report measurements of the dc voltage across Al/AlO$_x$/Ni$_{80}$Fe$_{20}$/Cu tunnel junctions with precessing magnetization of Ni$_{80}$Fe$_{20}$. The surprisingly large observed voltage is about μV, which seems to *qualitatively* agree with the spin pumping theory [5, 6] but requires an *unreasonably large* spin-mixing conductance of the *FM*/tunnel-barrier contact. We conclude that a new nonequilibrium phenomenon, which dynamically couples the spin and charge degrees of freedom, exists in tunneling structures. The rest of this Letter presents details of our experiment, and we conclude by speculating on possible theoretical scenarios responsible for these surprising experimental results.

The experimental setup is illustrated in Fig. 1. A tunnel junction was fabricated on the signal conductor of a coplanar waveguide (CPW) transmission line. The tunnel junction structure of Cu (100nm)/Al (10nm)/AlO$_x$ (2.3nm)/Ni$_{80}$Fe$_{20}$ (20nm)/Cu (70nm)/Au (25nm) was fabricated on a Si substrate with a 1 μm thick thermal oxide layer, by using magnetron sputtering deposition and conventional microfabrication processing. The bottom-most 100 nm Cu layer was patterned into the CPW designed to have 50 Ω characteristic impedance in the absence of the tunnel structure. The aluminum oxide tunnel barrier was formed by plasma oxidation. The size of the tunnel junction pillar was 50×50 μm$^2$, and the dc junction resistance was measured to be 67 kΩ. A microwave signal from a vector network analyzer (Agilent 8753B) was introduced through the CPW and generated a microwave magnetic field $H_{rf}$ that was linearly polarized in the plane of the tunnel junction. The external dc magnetic field (-120 Oe to 120 Oe) was swept along the axis of the CPW (the *y*-direction), so that the magnetization changed its direction within the *x-y* plane. The precessing spin mainly rotated around the *y*-axis.

Two electrical probe tips were used to measure the dc voltage across the junction. The microwave input signal was varied from 0.7 GHz to 4 GHz, with power up to 18 dBm amplitude-modulated with a 400 Hz sinusoidal signal to allow for lock-in detection. It should be mentioned that there was always a few tens of microvolt background voltage at the detector due to microwave noise. We found that the background voltage was directly proportional to the input microwave power. We can thus maintain the same power applied to the device at different frequencies by slightly tuning the nominal input power (±1 dBm) to maintain the same background voltage within 20% error. This makes it possible to compare the data between different frequencies without concerning the frequency dependence of the CPW impedance which changes the power input to the device slightly.

Figure 2 shows the dc voltage as a function of the external magnetic field in the Al/AlO$_x$/Ni$_{80}$Fe$_{20}$/Cu tunnel junction. At each microwave frequency, the voltage peaks of magnitude $\Delta V$ appear symmetrically at positive and negative fields. The peak field as a function of the microwave frequency shown in Fig. 3 agrees well with the values we obtained from the flip-chip CPW FMR measurements. The Kittel formula [12] fits the data with reasonable parameters, $4\pi M_s = 9 \text{ kG}$, $H_k = 19 \text{ Oe}$, and gyromagnetic ratio $\gamma = 0.0176 \text{ s}^{-1}\text{Oe}^{-1}$, confirming that the dc voltage peak appears at the uniform FMR mode of the Ni$_{80}$Fe$_{20}$ layer. The peak magnitude reaches about 1 μV at 2GHz which is much larger than the maximum value of about 250 nV at 14.5 GHz reported in Ref. [2] for a Pt/Ni$_{80}$Fe$_{20}$/Al structure. Figure 4 shows microwave power and frequency dependence of $\Delta V$, which increases with increasing microwave power [Fig. 4 (a)]. We also plot $\Delta V$ as a function of precession cone angle in Fig. 4(b). The precession cone angle of Ni$_{80}$Fe$_{20}$ was determined by the change in the tunnel resistance at the FMR field in IrMn/Fe$_{70}$Co$_{30}$/AlO$_x$/Ni$_{80}$Fe$_{20}$ magnetic tunnel junctions [13] with 20mV bias voltage so that the dc voltage effect of the microvolt order we are discussing here can be neglected.  A clear dip in antiparallel states and a peak in parallel states are observed corresponding to FMR fields, and the precession angle $\theta$ can be determined from $\Delta R/R \propto (1-\cos\theta)$.  At 10 dBm power input, the precession cone angle was around 17° [13]. As the applied frequency increased, $\Delta V$ increases almost linearly, as shown in Fig. 4(c).

Before attempting to interpret our results, we have to examine carefully the rectification effects which could induce similar dc voltage response. Possible rectification effect may arise from both the time-dependent anisotropy magnetoresistance (AMR) effect and the anomalous Hall effect (AHE) discussed in Refs.[14]. The current due to these two effects is given by

$$\mathbf{j}' = -\frac{\Delta\rho}{\rho M^2}(\mathbf{j}\cdot\mathbf{M})\mathbf{M} + R\sigma\mathbf{j}\times\mathbf{M}, \qquad (1)$$

where $\mathbf{j}$ is the current, $\sigma$ is the conductivity, $\rho$ is the resistivity, $\Delta\rho$ is the magnetoresistive anisotropy, and $R$ is the anomalous Hall constant. The magnetization precessing around the y-axis is described by the vector $\mathbf{M} = (M\sin\omega t\sin\theta, M\cos\theta, M\cos\omega t\sin\theta)$. The microwave-induced current across the junction along the microwave electric field direction (z-axis) is given by $\mathbf{j} = (0,0, j\cos(\omega t+\alpha))$, where $\alpha$ is the phase difference with respect to the phase of spin precession. The z-component of Eq. (1) is of interest to our experiments, $j'_z = -\Delta\rho/\rho\, j\cos(\omega t+\alpha)\cos^2\omega t\sin^2\theta$. The time average of $j'_z$ is zero, allowing us to conclude that there is no dc component in the z-direction in our sample. The result holds even if the precession axis fluctuates in the x-y plane. Thus, in our sample configuration we expect no dc voltage generated due to the rectification effect. We purposely broke the tunnel barrier to investigate Al/Ni$_{80}$Fe$_{20}$/Cu contact and also made Cu/AlO$_x$/Al/Ni$_{80}$Fe$_{20}$/Cu junctions. In both cases, no $\Delta V$ was observed within our measurement sensitivity of about 100 nV. This implies that the large $\Delta V$ in Al/AlO$_x$/Ni$_{80}$Fe$_{20}$/Cu was indeed developed due to the AlO$_x$/Ni$_{80}$Fe$_{20}$ interface.

Let us now try to interpret our results within the framework of the standard spin pumping theory [5-7]. At the FMR, a steady precession of the magnetization of the FM layer pumps a spin current into the adjacent NM according to [5, 7]

$$\mathbf{I}_s^{pump} = \frac{\hbar}{4\pi}\left(\operatorname{Re} g^{\uparrow\downarrow} \mathbf{m}\times\frac{d\mathbf{m}}{dt} + \operatorname{Im} g^{\uparrow\downarrow}\frac{d\mathbf{m}}{dt}\right), \qquad (2)$$

where $\mathbf{m}$ is the unit vector along the instantaneous direction of the precessing magnetization and $\operatorname{Re} g^{\uparrow\downarrow}$ ($\operatorname{Im} g^{\uparrow\downarrow}$) is the real (imaginary) part of the dimensionless interfacial spin-mixing conductance (in units of $e^2/h$) which describes spin transport perpendicular to $\mathbf{m}$ at the *FM/NM* interface [5, 15, 16]. For transparent intermetallic *FM/NM* contacts $\operatorname{Im} g^{\uparrow\downarrow}$ is typically neglected because of being much smaller than $\operatorname{Re} g^{\uparrow\downarrow}$ [5-7], while for low transparent contacts we find $\operatorname{Re} g^{\uparrow\downarrow}/g = 0.5$ and $\operatorname{Im} g^{\uparrow\downarrow}/g \simeq 0.5$ (using the simple Stoner model for *FM* and random binary alloy with a gap in the energy spectrum for the tunnel barrier as *NM* layer; $g$ is the total charge conductance of the junction). The possibility for non-negligible $\operatorname{Im} g^{\uparrow\downarrow}$ for tunneling interfaces is also highlighted by recent first principles calculations [17]. The injected spin current builds up a spin accumulation $\mu_s$ in the *NM* layer (close to the *FM/NM* interface) when the spin-flip relaxation rate in *NM* is smaller than the spin injection rate.

This, in turn, drives a backward flowing spin current $\mathbf{I}_s^{back}$ into the precessing *FM* [5]. The backward flowing spin current parallel to the magnetization can be absorbed by the FM, in the presence of spin-flip processes. Due to spin-dependent bulk and interface conductances, this absorbed spin current is converted into charge accumulation at the *FM/NM* interface [18]. The maximum value of the voltage drop $V_{dc}$ across the *FM/NM* interface, at fixed frequency $\omega$ and cone angle $\theta$ of magnetization precession (assuming the frequency is much greater than the characteristic spin-flip rate in the normal metal and the ferromagnet is thicker than its bulk spin-diffusion length), is obtained for *NM* layer thickness much smaller than its spin-diffusion length $d_N << \lambda_{ds}^N$ as [6]:

$$V_{dc} = \frac{p_\omega g_F \sin^2\theta \cos\theta}{(g_\omega - p_\omega^2 g_\omega + g_F)(\eta_N + \sin^2\theta) + (1-p_\omega^2)g_F \cos^2\theta\, g_\omega/g_\omega^{\uparrow\downarrow}} \frac{\hbar\omega}{2e}, \qquad (3)$$

where $\eta_N = g_N / g_\omega^{\uparrow\downarrow} \tanh(d_N/\lambda_{sd}^N)$ with the thickness $d_N$ and the spin diffusion $\lambda_{sd}^N$ of the NM layer, $g_F$ and $g_N$ parameterize the transport properties of the bulk *FM* and *NM*, $g_\omega^{\uparrow\downarrow}$ is the real part of the effective spin-mixing conductance, $p_\omega = \left(g_\omega^\uparrow - g_\omega^\downarrow\right)/\left(g_\omega^\uparrow + g_\omega^\downarrow\right)$ is the interfacial spin-polarization, $g_\omega = g_\omega^\uparrow + g_\omega^\downarrow$ is the sum of spin-up and spin-down effective conductances of the *FM/NM* interface. The "effective" interfacial transport quantities are in general frequency-dependent since they have to be evaluated for the interface resistance in series with a *NM* resistor of length $L_\omega = \sqrt{D_{NM}/\omega}$ over which the oscillating transverse components of spin accumulation in *NM* (with diffusion constant $D_{NM}$) are averaged to zero, although this is not important in practice for high-impedance tunnel barriers. The voltage drops $V_{dc}^{1,2}$ emerging at each of the two *FM/NM* contacts will differ from each other when conductances $g_\omega^{\uparrow\downarrow}$ and/or spin-flip diffusion lengths $\lambda_s^N$ on two sides of the multilayer are substantially different, as observed by measuring $\Delta V = V_{dc}^1 - V_{dc}^2$ on lateral Pt/Ni$_{80}$Fe$_{20}$/Al device in a recent experiment [2].

To compare the spin pumping theory with our results, we assume that spin-mixing conductance is governed by the AlO$_x$/Ni$_{80}$Fe$_{20}$ interface, but there is no spin flipping inside the barrier and spin accumulation is induced in the Al layer. Since the interface conductance $g_\omega$ for the low transparency tunnel barrier is much smaller than small $g_F$, the first term in the denominator of Eq. (3) reduces to $g_F \left( \eta_N + \sin^2 \theta \right)$, so that Eq. (3) can be parameterized with $g_\omega^{\uparrow\downarrow}/g_\omega$, $\eta_N$, $p_\omega$, and $\theta$. Using $p_\omega = 0.3$ for the AlO$_x$/Ni$_{80}$Fe$_{20}$ interface, Fig. 4(b) shows the best fitting (solid line) of our results by Eq. (3), where we extract $g_\omega^{\uparrow\downarrow}/g_\omega \approx 3.4$ and $\eta_N \approx 0$ from the fit. We found that $\eta_N$ has to be set to zero in order to fit our data, which requires that $g_N$ is roughly comparable or smaller than $g_\omega^{\uparrow\downarrow}$, while we expect the former to be several orders of magnitude larger than the latter, using the measured tunneling conductance. This is the first discrepancy between our results and an attempt to explain them using standard interfacial spin pumping theory originally developed [5, 7] and experimentally confirmed [2] for intermetallic *FM/NM* contacts. On the other hand, linear fitting of $\Delta V$ vs. frequency in Fig. 4(c) yields the slope of 0.85 µV/GHz, and $g_\omega^{\uparrow\downarrow}/g_\omega \approx 6.1$ at precession cone angle of 17° by Eq. (3). These values are larger than the typical value $g_\omega^{\uparrow\downarrow}/g_\omega \approx 1$ for transparent intermetallic contacts, which is *highly unexpected* when compared to standard estimates [16] of $g_\omega^{\uparrow\downarrow}/g_\omega$ for trivial (non-magnetic) tunnel contacts.

Voltage generation based on the spin pumping mechanism [6] is based on spin injection into the normal metal, across the *FM/NM* interface, with its subsequent diffusion, relaxation, and backflow into the ferromagnet, which is ultimately responsible for the build-up of the voltage drop across the contact. A tunnel barrier exponentially impedes electron flows (and thus spin currents) across the *FM/NM* contact, and one, therefore, would *not expect* a significant voltage generation by the spin-pumping mechanism. This is the reason why we were not able to reach a quantitative agreement with the theory. The tunnel barrier essentially cuts off the normal metal from the *FM*, while a voltage probe may now be thought of as a nonintrusive probe of dynamic processes within the *FM*. If the magnetization dynamics can generate nonequilibrium spin accumulation inside the ferromagnet, in analogy with the pumped spin generation in the normal metal (presumably requiring spin-orbit or other spin-flip processes in the *FM*), the voltage measured by the *FM* may in fact be probing this spin accumulation rather than a nonlocal spin pumping process. Exploring this possibility requires further theoretical analysis and other nonlocal probes of the magnetization dynamics. Finally, we note that our theoretical discussion completely disregarded many-body effects due to electron-electron interactions, which may modify substantially the predictions of the standard spin pumping theory, especially if we drive the magnetization dynamics beyond the linearized regime.

In conclusion, we observed a *large dc voltage*, of the order of microvolts, across the Al/AlO$_x$/Ni$_{80}$Fe$_{20}$/Cu tunnel junctions, due to a dynamic spin and charge coupling driven by the precessing magnetization of a single Ni$_{80}$Fe$_{20}$ ferromagnetic layer at ferromagnetic resonance. By short circuiting the tunnel barrier, we demonstrated that the observed dc voltage mainly arises from the Al/AlO$_x$/Ni$_{80}$Fe$_{20}$ contact. The phenomenon appears qualitatively similar to the predictions of the spin pumping formalism, but a quantitative analysis shows a number of discrepancies with the standard theory. This suggests a new nonequilibrium mechanism for the spin and charge coupling, which is responsible for the voltage generation much larger than that observed very recently for intermetallic interfaces [2]. We speculate on the role of intrinsic dynamic processes in the ferromagnet and the effects of the electron-electron interactions, as possible culprits for our observations, but a more thorough theoretical analysis is desirable in the future.

We thank M. D. Stiles and S.-T. Chui for illuminating discussions. This work was supported by NSF DMR Grant No. 0405136, and DOE DE-FG02-07ER46374.

**Figure captions:**

FIG. 1. (Color online) Schematic diagram of the sample structure (a) and the measurement geometry (b). The arrow in the $Ni_{80}Fe_{20}$ layer indicates the magnetization direction. An external dc field $H_{ex}$ is applied in the y-direction and the rf magnetic field $H_{rf}$ is applied along the x-direction. A coplanar microwave probe feeds microwave signals through the coplanar waveguide. DC voltage across the junction (TJ) is measured between top of the junction and signal line of the CPW.

FIG. 2. (Color online) The dc voltage $\Delta V$ generated across the $Al/AlO_x/Ni_{80}Fe_{20}/Cu$ tunnel junction as a function of the externally applied static magnetic field. The frequency of the applied rf field ranges from 1.8 to 2.8 GHz. The background voltage is subtracted for comparison purpose.

FIG. 3. (Color online) The frequency dependence of static magnetic field at which the dc voltage peak (circles) appears in Fig. 2. The crosses label the frequency dependence of the resonance field obtained from FMR measurements on flip-chip structures in the CPW line. The curve is a fit to the Kittel formula [12].

FIG. 4. The amplitude of the dc voltage $\Delta V$ measured across $Al/AlO_x/Ni_{80}Fe_{20}/Cu$ device as function of: (a) microwave power; (b) precession cone angle; and (c) microwave frequency at 10 dBm. Solid lines in (b) and (c) are the fit to Eq. (3) as described in the text.

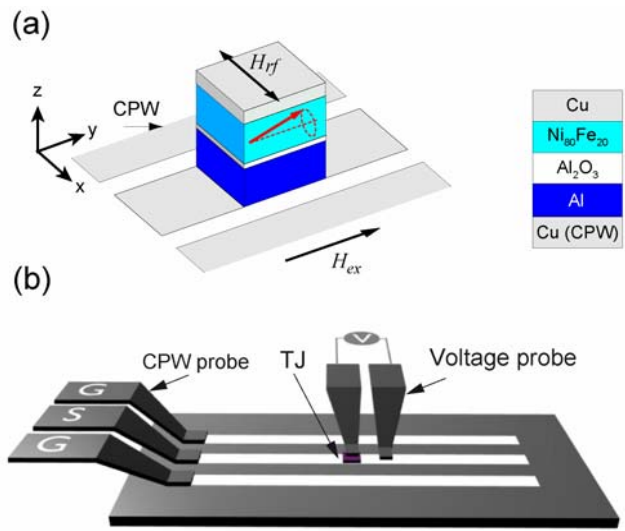

Fig.1 T. Moriyama et al.

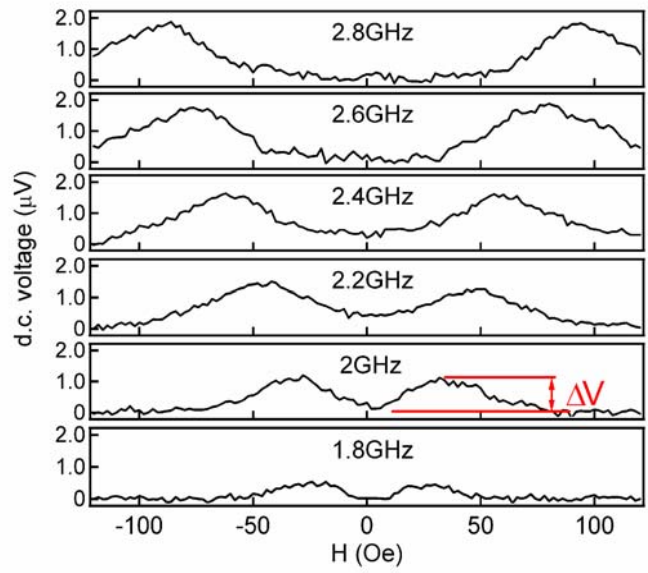

Fig. 2 T. Moriyama et al.

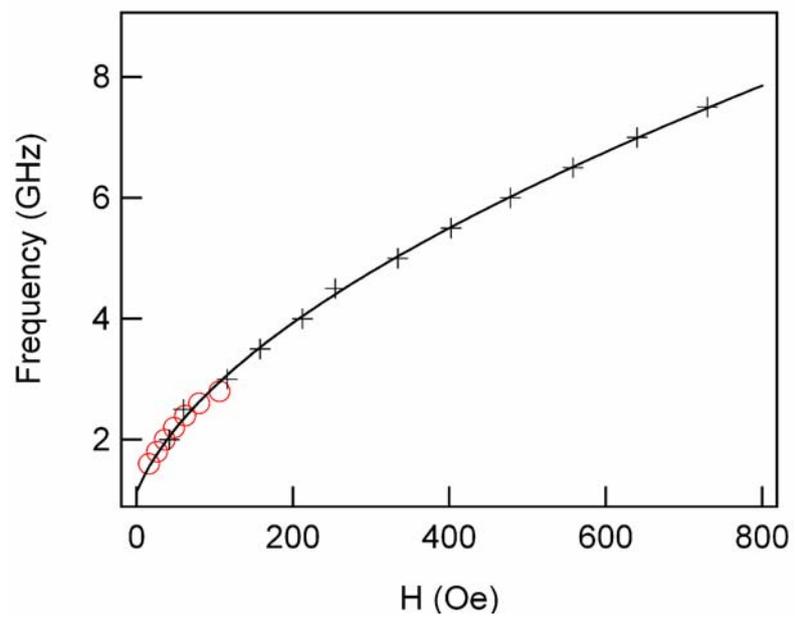

Fig.3 T. Moriyama et al.

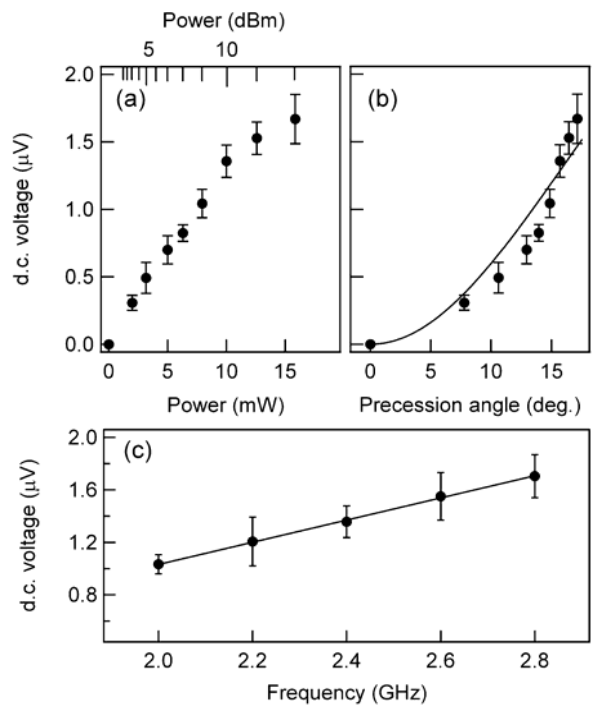

Fig. 4 T. Moriyama et al.